\documentclass[useAMS,usenatbib,usegraphicx]{mn2e}
\title[Infrared Ca\,{\normalsize \it II} triplet as metallicity indicator]{The Near Infrared Ca\,{\Large\bf II} Triplet as Metallicity Indicator: II Extension to extremely metal--poor metallicity regimes.\footnote{Based on observations collected at: the 2.5 Isaac Newton and 4.2 William Herschel Telescopes operated on the island of La Palma by the Isaac Newton Group in the Spanish Observatorio del Roque de los Muchachos of the Instituto de Astrofísica de Canarias; the Centro Astronómico Hispano Alem\'an (CAHA) at Calar Alto, operated jointly by the Max-Planck Institut für Astronomie and the Instituto de Astrof\'{\i}sica de Andaluc\'{\i}a (CSIC); Cerro Tololo Inter-American Observatory, National Optical Astronomy Observatory, which are operated by the Association of Universities for Research in Astronomy, under contract with the National Science Foundation; and the European Southern Observatory, Chile, within the observing programs 070.B-0398 and 074.B-0446. }}
\author[R. Carrera et al.]{R. Carrera$^{1,2}$\thanks{E-mail:
rcarrera@iac.es} E. Pancino$^{3,4}$ C. Gallart$^{1,2}$ A. del Pino$^{1,2}$\\
$^{1}$Instituto de Astrof\'{\i}sica de Canarias, La Laguna E-3200, Tenerife, Spain\\
$^{2}$Departamento de Astrof\'{\i}sica, Universidad de La Laguna, La Laguna E-38205, Tenerife, Spain\\
$^{3}$Osservatorio Astronomico di Bologna, via Ranzani 1, I-40127 Bologna, Italy\\
$^{4}$ASI Science Data Center, I-00044 Frascati, Italy}
\begin{document}

\date{Accepted XXX. Received XXX; in original form XXX}

\pagerange{\pageref{firstpage}--\pageref{lastpage}} \pubyear{2002}

\maketitle

\label{firstpage}

\begin{abstract}
We extend our previous calibration of the infrared \mbox{Ca\,{\sc ii}} triplet as metallicity indicator to the metal-poor regime by including observations of 55 field stars with [Fe/H] down to -4.0~dex. While we previously solved the saturation at high-metallicity using a combination of a Lorentzian plus a Gaussian to reproduce the line profiles, in this paper we address the non-linearity at low-metallicity following the suggestion of \citet{starkenburg2010} of adding two non-linear terms to the relation among the [Fe/H], luminosity, and strength of the Calcium triplet lines. Our calibration thus extends from -4.0 to +0.5 in metallicity and is presented using four different luminosity indicators: V-V$_{HB}$, M$_V$, M$_I$, and M$_K$. The calibration obtained in this paper results in a tight correlation between [Fe/H] abundances measured from high resolution spectra and [Fe/H] values derived from the CaT, over the whole metallicity range covered.
\end{abstract}


\section{Introduction}\label{sec1}

The infrared \mbox{Ca\,{\sc ii}} triplet (CaT) lines at 8498, 8542, and 8662 \AA, easily
differentiated even in low- and medium-resolution spectra (R$\sim$2000-10000), are the main features of the
near infrared spectra of late-type giant stars. Since their strengths change as a function of the metal content, they have been widely used as metallicity indicators in a variety of systems like open
\citep[e.g.][]{cole2004,warren_cole2009,carrera2012} and globular \citep[e.g.][]{armandroffzinn1988,rutledge1997b} clusters,
dwarf spheroidal \citep[e.g.][]{armandroff_dacosta1991,pont2004,battaglia2008} and irregular
\citep[e.g.][]{dacosta1998,tolstoy2001,carrera2008b,parisi2010} galaxies, and even in a more complex system like the Large Magellanic Cloud \citep[e.g][]{olszewski1991,cole2005,carrera2008a,carrera2011}.

Initially, the CaT lines were employed to determine metallicities in relatively metal-poor and old systems such as the Galactic globular clusters \citep[e.g.][]{armandroffzinn1988,armandroff_dacosta1991,rutledge1997b}.
Most of these works utilized the classical approach of determining the equivalent widths of the CaT lines by fitting their profiles with Gaussian functions. The direct integration is hampered by the contamination of the wings with weak neutral metal lines and molecular bands.
There were some attempts of extending this calibration to more metal-rich and
younger regimes. All of them were hindered by the non-Gaussian shape of damping wings in strong lines of stars with [Fe/H]$>$-1 \citep[e.g.][]{suntzeff1992,suntzeff1993,rutledge1997b}. To address this issue, \citet{cole2004} proposed to fit the line profile with a sum of a Gaussian and a Lorentzian functions. For weak lines the Lorentzian component is equal to zero and the line profile is reproduced by a single Gaussian. In the case of strong lines, the Gaussian reproduces the line core but the Lorentzian component is necessary to properly sample the wings. Using this procedure \citet[][hereafter Paper I]{carrera2007} extended the empirical
calibration of the strength of the CaT lines as metallicity indicator for
13$\leq$Age(Gyr)$\leq$0.25 and -2.2$\leq$[Fe/H]$\leq$+0.47. They also demonstrated that the age influence is negligible.

Curiously, none of the studies performed until that moment, and in particular those devoted to the metal-poor dwarf spheroidal galaxies, had detected stars more metal-poor than [Fe/H]$\leq$-3 \citep[e.g.][]{helmi2006}. In contrast, the halo of the Milky Way contains a significant
number of these objects which might reveal valuable information about the
chemical evolution history of a galaxy, as they represent the most pristine (and
probably thus the oldest) stars in the system. By comparison with metallicities derived from high-resolution spectroscopy analysis, \citet{battaglia2008} found that CaT calibrations saturate for [Fe/H]$\leq$-2.5, and therefore, they are not able to detect stars more metal-poor than this value.

\begin{table*}
 \centering
\caption{Cluster sample.\label{clustersample}}
  \begin{tabular}{@{}lcccccccc@{}}
  \hline
Cluster & [Fe/H] & [Ca/H] & Ref. & [Fe/H]$_{Paper I}$ &
Age(Gyr) & $(m-M)_V$ & $E(B-V)$ & Ref.\\
 \hline
NGC 104 (47 Tuc) & $-0.76\pm0.02$ & $-0.45\pm0.06$ & 1,23 & $-0.67\pm0.03$ & $10.7\pm1.0$ & 13.32 & 0.05 & 8,13 \\
NGC 188 & $-0.03\pm0.04$ & $-0.08\pm0.03$ & 2 & $-0.07\pm0.04$ & $6.30\pm0.3$ & 11.44 & 0.09 & 9,17 \\
NGC 288 & $-1.32\pm0.02$ & $-0.90\pm0.05$ & 1,23 & $-1.07\pm0.03$ & $11.3\pm1.1$ & 14.64 & 0.03 & 8,13 \\
NGC 362 & $-1.30\pm0.04$ & $-1.16\pm0.05$ & 1,28 & $-1.09\pm0.03$ & $8.7\pm1.5$ & 14.75 & 0.05 & 8,13 \\
NGC 1851 & $-1.18\pm0.08$ & $-0.85\pm0.03$ & 1,25 & & $9.2\pm1.5$ & 15.49 & 0.02 & 8,13 \\
Berkeley 17 & $-0.10\pm0.09$ & $-0.15\pm0.10$ & 3 & & $9\pm0.5$ & 14.27 & 0.61 & 27 \\
NGC 1904 (M79) & $-1.58\pm0.02$ & $-1.30\pm0.04$ & 1,23 & $-1.37\pm0.05$ & $11.7\pm1.3$ & 15.53 & 0.01 & 8,13 \\
Berkeley 20 & $-0.30\pm0.02$ & $-0.22\pm0.06$ & 3 & $-0.49\pm0.05$ & $4.05\pm0.7$ & 15.84 & 0.38 & 9,22 \\
NGC 2141 & $+0.00\pm0.16$ & $-0.11\pm0.11$ & 15 & $-0.14\pm0.05$ & $2.45\pm0.9$ & 14.15 & 0.40 & 9,14 \\
Collinder 110  & $+0.03\pm0.02$ & $-0.04\pm0.02$ & 4 &  & $1.3\pm0.2$ & 13.04 & 0.40 & 24 \\
NGC 2298 & $-1.96\pm0.04$ & $-1.52\pm0.04$ & 1,29 & $-1.74\pm0.04$ & $12.6\pm1.4$ & 15.54 & 0.13 & 8,13 \\
Melote 66 & $-0.33\pm0.03$ & $-0.22\pm0.05$ & 16 & $-0.38\pm0.06$ & $5.3\pm1.4$ & 13.63 & 0.14 & 9,26 \\
Berkeley 39 & $-0.21\pm0.01$ & $-0.22\pm0.07$ & 19 & & $7.0\pm1.0$ & 13.24 & 0.11 & 9,26 \\
NGC 2682 (M 67) & $+0.05\pm0.02$ & $-0.11\pm0.03$ & 4 & $-0.03\pm0.03$ & $4.3\pm0.5$ & 9.65 & 0.04 & 9,17 \\
NGC 3201 & $-1.51\pm0.02$ & $-1.21\pm0.07$ & 1,23 & $-1.24\pm0.12$ & $11.3\pm1.1$ & 14.17 & 0.21 & 8,13 \\
NGC 4590 (M 68) & $-2.27\pm0.04$ & $-2.00\pm0.04$ & 1,23 & $-2.00\pm0.03$ & $11.2\pm0.9$ & 15.14 & 0.04 & 8,13 \\
NGC 5927 & $-0.29\pm0.07$ &  & 1 & & $10.9\pm2.2$ & 15.81 & 0.47 & 30,13 \\
NGC 6352 & $-0.62\pm0.05$ & $-0.36\pm0.07$ & 1,31 & $-0.64\pm0.02$ & $9.9\pm1.4$ & 14.39 & 0.21 & 8,13 \\
NGC 6528 & $+0.07\pm0.08$ & $+0.30\pm0.08$ & 1,32 & $-0.17\pm0.02$ & $11.2\pm2.0$ & 16.16 & 0.55 & 10,18 \\
NGC 6681 (M 70) & $-1.62\pm0.08$ &  & 1 & $-1.35\pm0.03$ & $11.5\pm1.4$ & 14.93 & 0.07 & 8,13 \\
NGC 6705 (M 11) & $+0.10\pm0.07$ & $-0.09\pm0.06$ & 5 & $+0.07\pm0.05$ & $0.25\pm0.1$ & 12.88 & 0.43 & 11 \\
NGC 6791 & $+0.47\pm0.07$ & $+0.32\pm0.08$ & 6 & $+0.47\pm0.04$ & $12.0\pm1.0$ & 13.07 & 0.09 & 12 \\
NGC 6819 & $+0.09\pm0.03$ & $+0.05\pm0.06$ & 7 & $+0.07\pm0.03$ & $2.9\pm0.7$ & 12.35 & 0.14 & 9,20,7\\
NGC 7078 (M 15) & $-2.33\pm0.02$ & $-2.07\pm0.10$ & 1,23 & $-2.12\pm0.04$ & $11.7\pm0.8$ & 15.31 & 0.09 & 8,13\\
NGC 7789 & $+0.04\pm0.07$ & $-0.14\pm0.09$ & 4 & $-0.04\pm0.05$ & $1.3\pm0.3$ & 12.20 & 0.28 & 9,4,21\\
\hline
\end{tabular}
\begin{minipage}{170mm}
References: (1) \citet{carretta2009}; (2) \citet{jacobson2011}; (3) \citet{friel2005}
(4) \citet{pancino2010}; (5) \citet{gonzalez2000}; (6) \citet{carretta2007};
(7) \citet{bragaglia2001}; (8) \citet{salaris2002}; (9) \citet{salaris2004};
(10) \citet{feltzing2002}; (11) \citet{sung1999}; (12) \citet{stetson2003};
(13) \citet{rosenberg1999}; (14) \citet{carraro2001}; (15) \citet{jacobson2009};
(16) \citet{sestito2008}; (17) \citet{sarajedini1999}; (18) \citet{ortolani1992};
(19) \citet{friel2010}; (20) \citet{rosvick1998}; (21) \citet{gim1998};
(22) \citet{yong2005}; (23) \citet{carretta2010b}; (24) \citet{bragaglia2003}; 
(25) \citet{carretta2011}; (26) \citet{kassis1997}; (27) \citet{bragaglia2006}
(28) \citet{shetrone2000}; (29) \citet{mcwilliam1992}; (30) \citet{fullton1996}; (31) \citet{feltzing2009}; (32) \citet{carretta2001}.
\end{minipage}
\end{table*}

It is well known that the strength of the CaT lines does not
only depend on the metallicity but also on the temperature and gravity on the stellar surface. These contributions are removed using the fact that the strength of the CaT lines is linearly correlated with a luminosity indicator, used to trace the temperature and gravity variations, for stars in a range of 2-3 mag below the tip of the RGB for the metallicity range covered by open and globular clusters \citep[e.g.][Paper I]{armandroff_dacosta1991,rutledge1997a,cole2004}. The calibration of the CaT lines as metallicity indicator is defined by the linear relation between the zero-point of these sequences and the metallicity. In Paper I it was noticed that the relation between the strength of the CaT lines and luminosity is not exactly linear if a larger luminosity range below the tip of the RGB is sampled. Recently, using synthetic spectra, \citet{starkenburg2010}  showed that, together with the fact that these sequences are not linear, the CaT saturation at lower metallicities is also due to the assumption of a linear relation between the zero-points of these sequences and metallicity. They obtained a new calibration with two additional terms to account for these issues. 

However, this new calibration is based on synthetic spectra which suffer from uncertainties in
the physics used to obtain them, particularly in this extremely metal-poor regime. The goal of this paper is to study the behaviour of the CaT lines in extremely metal-poor stars ([Fe/H]$\leq$-2.5) and obtain a new calibration of the strength of the CaT lines as metallicity indicator valid for the widest range of metallicities from observed spectra. To do that, the cluster sample used in Paper I has been complemented with observations of metal-poor field stars. Both samples are described in Section~\ref{sec2}. The different luminosity indicators used are discussed in Section~\ref{sec3}. The reference metallicities used are explained in Section~\ref{sec4}. Section~\ref{sec5} accounts for the determination of the CaT lines equivalent widths and the CaT index definition. The new calibration of the CaT lines
as metallicity indicator is obtained in Section~\ref{sec6}, where it is also compared with other calibrations available in the literature. Finally, the main results of this paper are summarized in Section~\ref{sec7}.

\section{Observational material}\label{sec2}

\subsection{Cluster sample}

The CaT calibration obtained in paper I was obtained from observations of almost 500 RGB stars in 29 Galactic open and globular clusters, covering a total metallicity range of -2.33$\leq$[Fe/H]$\leq$+0.47. We used the same sample in this paper, so we refer the reader to Paper I for a detailed discussion about the observations and data
reduction of these stars. As in Paper I, Berkeley~32, NGC~2420, and NGC~2506 were excluded from the analysis because of the small number of stars ($\leq5$) observed in each of them. Moreover, we excluded the globular
cluster NGC~6715 (M54) with a well known intrinsic metallicity dispersion of $\sim$0.19 dex
\citep[e.g.][]{carretta2010}. We have complemented the sample with observations of stars in the old and metal-rich open cluster Berkeley~17. This cluster was observed in November, 2005 with AF2/WYFFOS at the William Herschel Telescope (WHT) at the Roque de los Muchachos Observatory (La Palma, Spain) with the same setup used in Paper I (see table~2 of Paper I). The data reduction was performed and the radial velocities were calculated using
IRAF\footnote{Image Reduction and Analysis Facility, IRAF is distributed by the
National Optical Astronomy Observatories, which are operated
by the Association of Universities for Research in Astronomy, Inc., under
cooperative agreement with the National Science Foundation.} packages following the same procedure described in Paper I. Basically, after bias, overscan subtraction, and trimming, with \textit{ccdproc},  dofibers were used to trace the apertures, make the flat-field correction, and perform the wavelength calibration. Finally, the sky lines were subtracted with a custom program (see Paper I for details). Eight of the observed stars have been confirmed as Berkeley~17
members from their radial velocity. From these stars we obtained a 
mean radial velocity of $V_r$=-79$\pm$11 km s$^{-1}$ in good agreement, within the
uncertainties, with other values available in the literature: -84$\pm$11 km s$^{-1}$ \citep{friel2002}, and -73.7$\pm$0.8 km s$^{-1}$ \citep{friel2005}. The final sample includes stars of 25 clusters which are listed in Table~\ref{clustersample} together with the main characteristics and the reference metallicities for each of them. 

\subsection{Metal-poor stars}
\begin{table*}
 \centering
\caption{Metal-poor star sample. The full version of this table is available in the online journal and in the CDS.\label{starsample}} 
 \begin{tabular}{@{}lccccccccccc@{}}
\hline
Star & T$_{eff}$  & log~g & [Fe/H] & [Ca/H] & Ref. & V & M$_V$ &
M$_I$ & M$_K$ & V$_r$ (km s$^{-1}$) & S/N \\
  \hline
CS31082-0001 & 4922$\pm$100 & 1.9$\pm$0.3 & -2.78$\pm$0.19 & -2.62$\pm$0.24 & 1 & 11.32 & -0.51$\pm$0.13 & -1.48$\pm$0.14 & -2.71$\pm$0.16 & 139$\pm$3 & 73.0 \\
CS22175-0007 & 5108$\pm$100 & 2.5$\pm$0.4 & -2.81$\pm$0.18 & -2.50$\pm$0.23 & 1 & 13.49 &  0.64$\pm$0.13 & -0.25$\pm$0.13 & -1.42$\pm$0.14 & -5$\pm$5 & 55.0 \\
HD15656 & 3990$\pm$100 & 1.8$\pm$0.3 & -0.16$\pm$0.17 & -0.25$\pm$0.20 & 6 & 5.16 & 0.18$\pm$0.10 & -1.25$\pm$0.10 &  -2.98$\pm$0.09 & -47$\pm$6 & 140.0 \\
\hline
\end{tabular}
\begin{minipage}{180mm}
References: (1) \citet{barklem2005}; (2) \citet{andrievsky2011}; (3) \citet{lai2008}
(4) \citet{johnson2002}; (5) \citet{mcwilliam1990}; (6) \citet{hollek2011} ;
(7) \citet{fulbright2000}; (8) \citet{wu2011}; (9) \citet{luck1985};
(10) \citet{pilachowski1996}; (11) \citet{mcwilliam1995}; (12) \citet{giridhar2001};
(13) \citet{zhang2009}.
\end{minipage}
\end{table*}

\begin{figure}
\includegraphics[width=84mm]{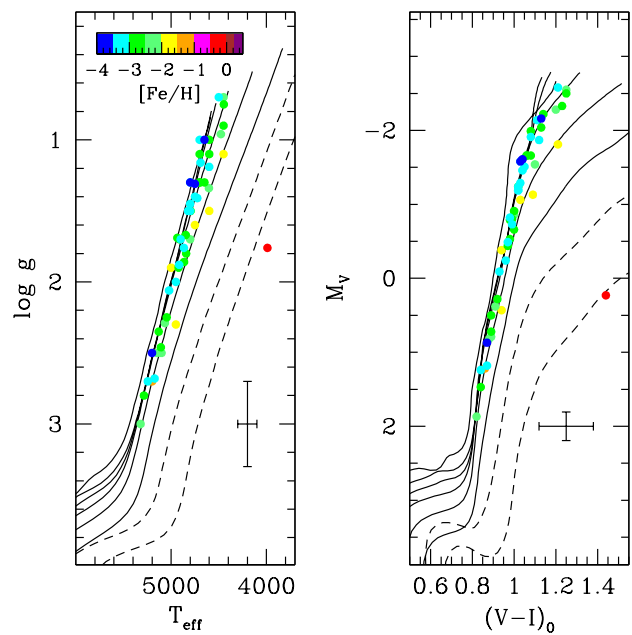}
\caption{Location of the observed metal-poor stars in the T$_{eff}$-log~g (left) and (V-I)$_o$-M$_V$ (right) planes. Each star has been coloured as a function of its metallicity. Overplotted are the Y$^2$ (Yale-Yonsei) isochrones: solid lines are 10 Gyr old isochrones of [$\alpha$/Fe]=+0.6 and -4.0$\leq$[Fe/H]$\leq$-1.5 with a step of 0.5 dex; dashed lines are 8 Gyr old isochrones with [$\alpha$/Fe]=+0.0 and [Fe/H]=-0.5 and 0.0. Typical errorbars are plotted in bottom-right corner of each panel.}
\label{fig_hr_stars}
\end{figure}

Unfortunately there are no clusters more metal-poor than [Fe/H]$\sim$-2.3
dex. Therefore, to investigate the behavior of the CaT lines at lower
metallicity we have complemented the cluster sample with observations of metal-poor field stars. These stars were
observed in two runs in April 2011 and
February 2012 with IDS at the Issac Newton Telescope (INT) located at the Roque
de los Muchachos Observatory (La Palma, Spain). The instrument
configuration used is described in detail in \citet{carrera2012}. The exposure times
were selected as a function of the magnitude of the stars in order to ensure a signal-to-noise ratio (S/N) greater
than 20. Observed stars, together with the S/N, and magnitudes in $V$ bandpass are listed in Table \ref{starsample}. The data reduction was performed in the same way as in Paper I for the cluster stars using IRAF packages. Several cluster stars previously studied in Paper I and a field metal-rich object, HD15656 ([Fe/H]=-0.16), were also observed in
both runs to ensure the homogeneity of the sample. The differences between the
equivalent widths obtained in each run are $\sim$0.1 \AA, which is similar to the
mean uncertainty in the equivalent width determination itself (see Section~\ref{sec5}). The radial velocity of each star was calculated using the fxcor task in IRAF using as template some of the stars with the best S/N in our sample. The final radial velocity for each target star was obtained as the average of the velocities obtained for each template, weighted by the width of the correlation peaks. The measured radial velocity for each star are listed in Table~\ref{starsample}.

\section{Luminosity Indicator}\label{sec3}

\begin{figure}
\includegraphics[width=84mm]{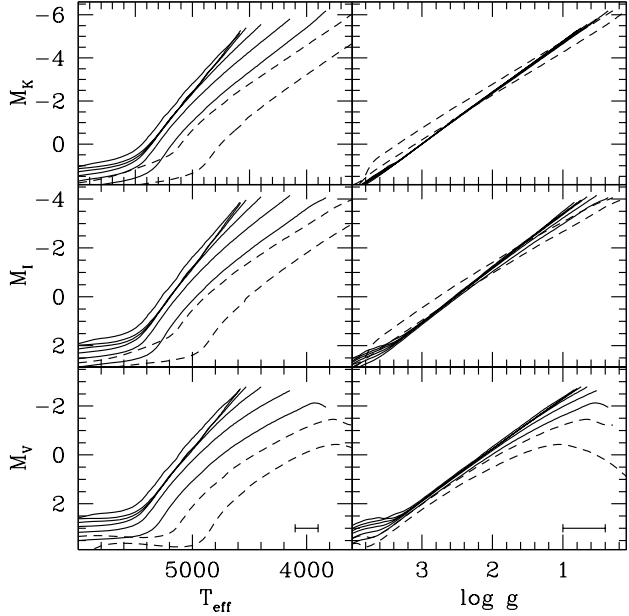}
\caption{Behaviour of T$_{eff}$ (left) and log~g (right) as a function of M$_V$ (bottom), M$_I$ (middle), and M$_K$ (top) for the same isochrones used in Fig.~\ref{fig_hr_stars}.}
\label{fig_isochrones}
\end{figure}

Several luminosity indicators have been used in the literature to remove the temperature and gravity contributions and leave only the abundance dependence in the CaT line strengths. The magnitude of a star relative to the position of the horizontal branch (HB) in the  $V$ filter, denoted as $V-V_{HB}$, is the most used one \citep[e.g.][]{armandroff_dacosta1991,rutledge1997b,cole2004}. Although it also removes any dependence on distance and reddening, it is hampered by the difficulty of defining the HB position in field stars, poorly populated clusters or galaxies with extended star formation histories. While in the case of old metal-poor systems, like globular clusters, the magnitude of the HB is defined as the RR Lyrae instability strip position, the red clump (RC) location has been used in the case of young metal-rich objects like open clusters. Some authors tried to account for this difference \citep[e.g.][]{dacosta1998} whereas others omit it \citep[e.g.][]{cole2004}. In the case of galaxies it is not possible to do this correction without an initial indication of the age of the stars. For example in a galaxy with multiple star formation epochs like the Large Magellanic Cloud, the difference between the position of the RR Lyrae instability strip and the the median magnitude of the RC can be as large as 0.4 mag. Since it is not possible to estimate the age of a given star prior to the measurement of its metallicity, this implies a metallicity uncertainty as large as $\sim$0.15 dex.

Alternatively, other authors used the absolute magnitude in the $V$ and/or $I$ Johnson-Cousins bandpasses  \citep[e.g.][]{pont2004,carrera2008a,carrera2008b,carrera2011} . Moreover in the case of external galaxies, the absolute magnitudes, which are obtained from distance and reddening, are often known with an accuracy better than 0.4 mag. In any case, determining stellar metallicities from CaT lines in the closed galaxies is usually hindered by the need of homogeneous photometry samples covering wide areas. Recently, \citet{warren_cole2009} and \citet{saviane2012} proposed to use the absolute magnitude in the $K_S$ bandpass. The Two Micron All-Sky Survey \citep[2MASS;][]{skrutskie2006} provides homogeneous photometry in near-infrared $J$, $H$, and $K_S$ bandpasses for almost the whole celestial sphere with an accuracy better than 0.03 mag and an astrometric precision of about 0.1 arcsec. Since the magnitude limit of 2MASS in $K_S$ band is 14.3, this implies that this survey has sampled the brightest RGB stars in almost all satellites of the Milky Way including the Magellanic Clouds. In order to obtain a  calibration as general as possible, we have used as luminosity indicators both the absolute magnitude in the $V$, $I$, and $K_S$ bandpasses and the magnitude relative to the HB position in the $V$ filter.

The absolute and relative magnitudes of cluster stars have been obtained from the distance modulus, reddening, and HB positions listed in Table~\ref{clustersample}. The reference photometry in $V$ and $I$ for each cluster was described in Paper I. In the case of Berkeley~17 the magnitudes of observed stars have been obtained from \citet{bragaglia2006}. The HB position of globular clusters were obtained from the updated version of the Harris globular cluster database\footnote{http://physwww.physics.mcmaster.ca/~harris/mwgc.dat} \citep{harris1996}. In the case of open clusters, the HB/RC position has been obtained from the original references of the photometry of each cluster (see Paper I) and from \citet{cole2004}.

Unfortunately, the absolute magnitudes of the field stars in our sample are unknown. To derive them we followed a similar procedure to that described by \citet{breddels2010}. It consists on calculating the absolute magnitudes from observable quantities using theoretical stellar evolution models. In our case, we used as input observable parameters the effective temperature (T$_{eff}$), surface gravity (log~g), and metallicity ([Fe/H]), derived from high-resolution spectroscopy. Because they properly sample the extremely metal-poor regimes of our field star sample we assumed the Y$^2$ Yonsei-Yale stellar evolution models  \citep{demarque2004} available at http://www.astro.yale.edu/demarque/yyiso.html. Using the \textit{YYmix2} interpolation routine provided together with the Y$^2$ models we created a grid of isochrones for -4.0$\leq$[Fe/H]$\leq$0.0 with a separation of 0.5 dex, 2$\leq$Age$\leq$12 Gyr with a step of 2 Gyr, and three different $\alpha$-elements abundances: [$\alpha$/Fe]=0.0, +0.3, and +0.6 dex.

In the left panel of Fig.~\ref{fig_hr_stars} a sample of isochrones of different metallicities have been plotted in the T$_{eff}$-log~g plane. The field stars in our sample have been overplotted with different colours, as a function of their metallicities. Since isochrones for certain ranges of ages and metallicities overlap it is not possible to infer unique age and metallicity, and therefore, they may imply uncertainties in the derived absolute magnitudes. For this reason we use a statistical approach which provides a probability distribution for each recovered magnitude.

In addition, the uncertainties of the input parameters affect the derived magnitudes. Typical errorbars for T$_{eff}$ and log~g are also plotted in bottom-right corner of left panel in Fig.~\ref{fig_hr_stars}. \citet{breddels2010} presented a detailed discussion about how T$_{eff}$ and log~g uncertainties affect the recovered absolute magnitudes for both main-sequence and RGB stars. In our case we limited our discussion to RGB objects. In Fig.~\ref{fig_isochrones} we have plotted the same isochrones as in Fig.~\ref{fig_hr_stars} illustrating the relations between T$_{eff}$ and log~g with M$_V$, M$_I$, and M$_K$ to investigate the impact of the uncertainties in the derived absolute magnitudes. It is clear that for RGB stars the absolute magnitudes are better constrained by log~g than by T$_{eff}$. Therefore a large log~g error implies a large uncertainty in the determination of M$_V$, M$_I$, and M$_K$. For example, a $\sigma_{log~g}$ of about $\pm$0.3 dex for a star with log~g=1.5 dex implies uncertainties of $\pm$0.5,  0.6, and 0.65 mag for M$_V$, M$_I$, and M$_K$, respectively using a 8 Gyr old isochrone with [Fe/H]=-2.0 dex, and [$\alpha$/Fe]=+0.6 dex. In the same way, a $\sigma_{T_{eff}}$ of about $\pm$100 k for a star of T$_{eff}$=4500 k implies uncertainties of $\pm$0.3,  0.35, and 0.4 mag for M$_V$, M$_I$, and M$_K$, respectively. Assuming different ages produces similar uncertainties as changing the temperature. Lower uncertainties are produced by the errors in [Fe/H] and [$\alpha$/Fe] ratios as was demonstrated by \citet{breddels2010}. For this reason, the [$\alpha$/Fe] ratio was not used as input parameter.

In order to match the observational input parameters with the absolute magnitudes provided by stellar evolution models we perform a classical $\chi^2$ minimisation defined as:

\begin{equation}\label{eq_chi}
 \chi^2=\sum_{i=0}^{n=3}\frac{(A_i-A_{i,model})^2}{\sigma_{A_i}^2}
\end{equation}

where A$_i$ are the three observable atmosphere parameters and A$_{i,model}$ the corresponding parameters of the model, as given by the set of isochrones. $\sigma_{A_i}$ represent the uncertainties of the observational quantities used as input. Although we used the specific uncertainties for each star, on average $\sigma_{T_{eff}}$ is $\sim$100 k, $\sigma_{log~g}$ is $\sim$0.25 dex, and $\sigma_{[Fe/H]}$ is $\sim$0.1 dex.

To take into account the impact of the input parameter uncertainties on the derived magnitudes, we performed a Monte Carlo realisation. For each star we computed 5000 fakes stochastically varying the input parameters, assuming that each of them behaves as a Gaussian probability distribution which mean and sigma are the given value and its uncertainty. Each fake is compared with each model point in the isochrone grid using Equation~\ref{eq_chi}. Instead of considering only the best solution, we adopted as absolute magnitudes for each fake the average of the values provided by the 20 best matches since different combinations of the input parameters can produce very similar $\chi^2$ values. The absolute magnitudes of the 5000 fakes for each star are Gaussianly distributed and therefore, we adopted as the absolute magnitude and its uncertainty the mean and sigma of each distribution. The obtained values and their uncertainties are listed in Table~\ref{starsample} and have been plotted in the right panel of Fig.~\ref{fig_hr_stars}. In general, the obtained magnitudes have uncertainties lower $\sim$0.2 mag.

In order to check the reliability of this method we have applied it to stars in four clusters: NGC~104, NGC~4590, NGC~7078, and Berkeley~39. They have been selected to sample a wide range of metallicities and in particular the metallicity range in which both field and cluster samples overlap around [Fe/H]$\sim$-2 dex. Therefore, this test also is valid to ensure that there is no bias between the two samples. For the three globular clusters we consider the red giant stars observed with high-resolution spectroscopy by \citet{carretta2009b} who determined temperatures, surface gravities, and metallicities for more than 50 stars in each system. In the same way, we use the red giant stars analysed by \citet{bragaglia2012} in the open cluster Berkeley~39.  The $I$ magnitudes of the globular clusters stars have been selected from the SUMO project \citep{monelli2013} since \citet{carretta2009b} do not provide them. We have used the method described above to estimate the absolute magnitudes from the temperature, surface gravity and metallicity. Obtained values have been compared with the absolute magnitudes derived from the distance modulus and reddening listed in Table~\ref{clustersample} as was done for cluster stars. We find that the differences between the values derived from the two procedures are on average -0.11$\pm$0.09, -0.12$\pm$0.09, and -0.12$\pm$0.12 for M$_V$, M$_I$, and M$_K$, respectively. These values are of the same order of the typical uncertainties of the absolute magnitudes derived from the method used for field stars. We conclude that our procedure provides reliable magnitudes within the uncertainties. This result also ensures that there is no significant bias between the absolute magnitudes derived for cluster and field stars.

Finally, to use $V-V_{HB}$ as luminosity indicator we have to define the HB position for each field star. Unfortunately, there is no much information about the position of the HB at extremely metal-poor regimes. \citet{starkenburg2010} estimated the position of HB from the empirical M$_V$-[Fe/H] relationship ($M_{V,HB}=0.23\times[Fe/H]+0.931$) obtained by \citet{catelan2008}. However, this relationship was obtained in the metallicity range covered by Galactic globular clusters and therefore, for [Fe/H]$\geq$-2.25 \citep[see][and references therein]{catelan2008}. To our knowledge, there is neither theoretical nor empirical better estimations of the position of the HB for more more metal-poor regimes. For this reason, we also used this relationship to estimate the position of the HB for our field star sample although its extrapolation to these regimes should be taken with care.

\section{Reference metallicities}\label{sec4}

Another important point to obtain the calibration of the CaT lines as metallicity indicator is the reference metallicity scale used. Unfortunately, there is no scale which includes clusters, both open and globular, and field stars. In fact there is even no common metallicity scale for open and globular clusters. For this reason the reference metallicities have been chosen from different sources as is explained below. Any attempt to homogenize them is clearly beyond the scope of this paper.

In the literature, we can find three metallicity scales for globular clusters. The traditional \citet{zinn1984} metallicity scale derived from low-resolution spectra and two more obtained from high-resolution spectra (R$>$20000): \citet{kraft2003} and \citet{carretta2009}, which is the updated version of \citet{carretta1997}. There are systematic differences among these three scales due to the resolution of the spectra, lines used for the analysis, etc. The different authors offer comparisons among them that can be used to switch from one scale to another. Because of the large number of clusters studied and the number of stars analysed in each of them we have chosen the \citet{carretta2009} metallicities as the reference for the globular clusters in our sample. It is also important that in this work, the [Fe/H] abundances were obtained using both \mbox{Fe\,{\sc i}} and \mbox{Fe\,{\sc ii}} spectral lines.

In the case of open clusters, several teams are working in deriving metallicities, and abundances of other chemical species, in an homogeneous way. Unfortunately, none of them include all the systems in our sample \citep[see][for a recent compilation of open clusters metallicities]{carrerapancino2011}. For this reason, we have chosen the reference metallicities of open clusters from different sources, with the constraint that they have to be derived from spectra of resolution equal or larger than $20000$ and using both \mbox{Fe\,{\sc i}} and \mbox{Fe\,{\sc ii}} spectral lines. These criteria have been also used to select the reference metallicities of the field stars in our sample. Moreover,  the average metallicity of each open cluster must have been obtained from at least three stars. The reference metallicities used for cluster and field stars are listed in Tables~\ref{clustersample} and \ref{starsample}, respectively.

\section{The CaT index}\label{sec5}
\begin{figure}
\includegraphics[width=84mm]{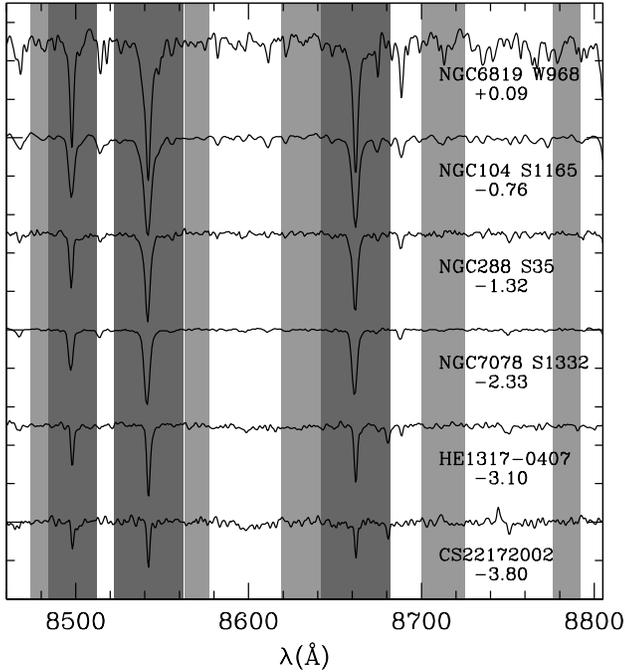}
\caption{Continuum (clear) and line (dark) bandpasses defined by
\citet{cenarro2001} overplotted onto stars of different metallicities.}
\label{fig_bandpasses}
\end{figure}

\begin{figure}
\includegraphics[width=84mm]{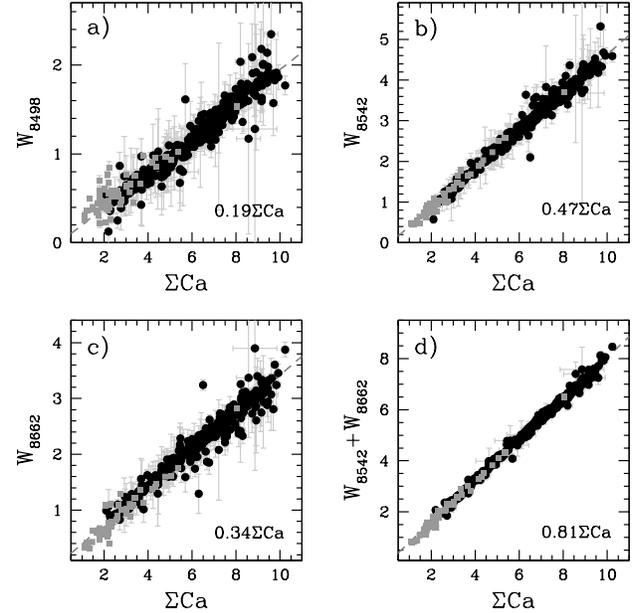}
\caption{Run of the strength of each individual CaT line versus the $\Sigma Ca$ index for both clusters (black) and metal-poor stars (grey), obtained as the sum of the equivalent widths of the three lines. The run of the sum of the equivalent widths of the two strongest lines versus $\Sigma Ca$ is also plotted. The fractional contribution of each line or combination, to the total $\Sigma Ca$ is labelled.}
\label{fig_comp}
\end{figure}

The strength of each CaT line was obtained in the classical way by determining the area between the spectral line inside a bandpass covering the feature, and the continuum level calculated in
several bandpasses among the three CaT lines. Although several definitions of these bandpasses can be found in the
literature (see Paper I for a comparison), in Paper I we selected those defined by \citet{cenarro2001}.
The line bandpasses are defined to cover completely each spectral line and in particular the wings of strong metal-rich lines. The continuum bandpasses are selected to avoid the presence of other spectral lines and
molecular bands. In Fig.~\ref{fig_bandpasses} we have overplotted the \citet{cenarro2001} bandpasses
onto stars of different metallicities. The selected bandpasses sample properly both metal-poor weak and metal-rich strong lines. Moreover, continuum bandpasses are not strongly affected by molecular bands in both metal-poor and metal-rich stars.
 
\begin{table}
 \centering
\caption{Equivalent widths of the three CaT lines for both cluster and field stars in our sample. The full version of this table is available in the online journal and in the CDS.\label{cluster_star_ew}}
  \begin{tabular}{@{}lccc@{}}
\hline
ID & $W_{8998}$ & $W_{8542}$ & $W_{8662}$ \\
\hline
NGC~104~L2705 & 1.33$\pm$0.02 & 3.21$\pm$0.03 & 2.73$\pm$0.03 \\ 
NGC~104~L2707 & 1.33$\pm$0.07 & 3.17$\pm$0.10 & 2.45$\pm$0.09 \\ 
NGC~104~L2720 & 1.25$\pm$0.12 & 2.96$\pm$0.13 & 2.30$\pm$0.12 \\ 
BD+053098 & 0.52$\pm$0.02 & 1.21$\pm$0.02 & 0.82$\pm$0.02\\ 
BD-185550 & 0.46$\pm$0.02 & 0.78$\pm$0.03 & 0.61$\pm$0.03\\ 
BD+233130 & 0.40$\pm$0.03 & 0.77$\pm$0.04 & 1.09$\pm$0.03\\
\hline
\end{tabular}
\end{table}

\begin{figure*}
\includegraphics[width=150mm]{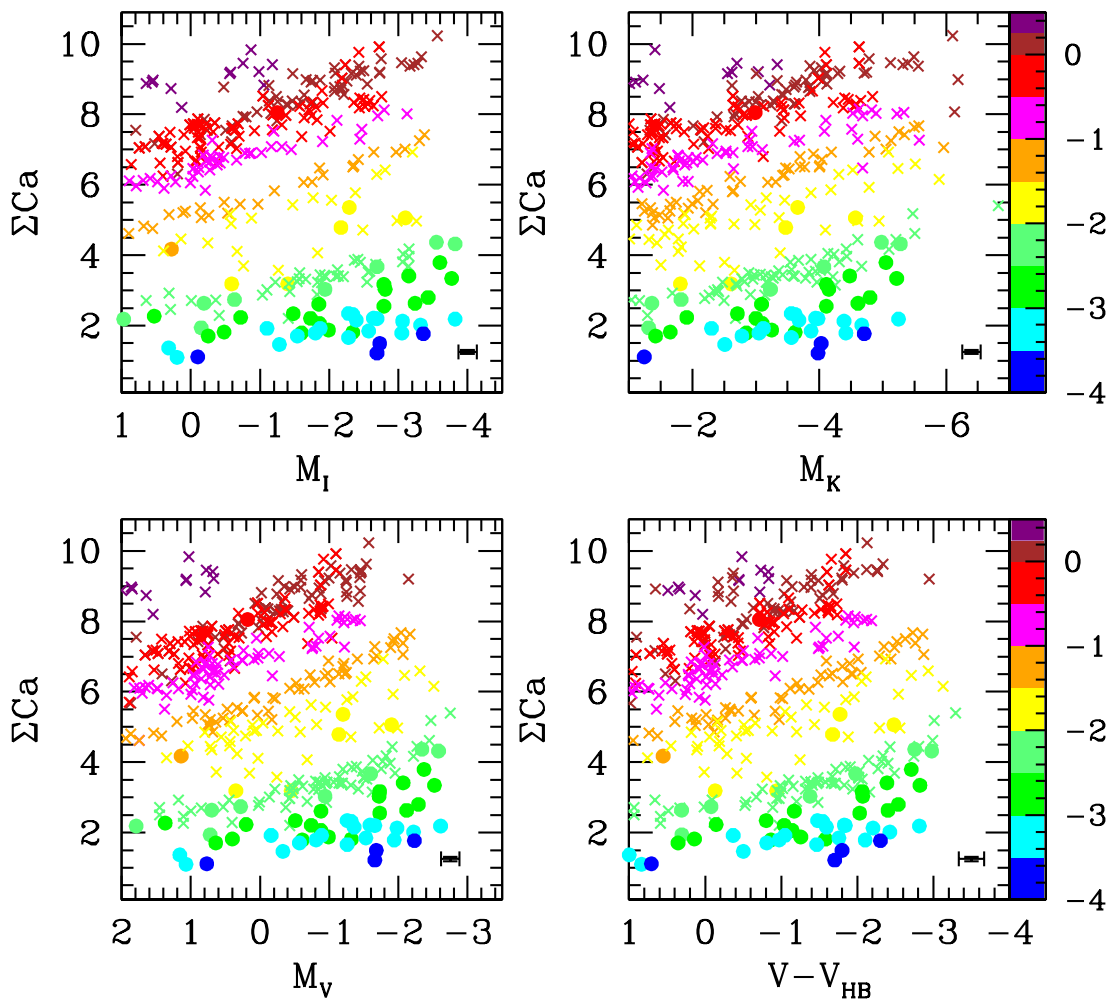}
\caption{Run of $\Sigma Ca$ as a function of $M_V$, $V-V_{HB}$, $M_I$, and $M_K$. Stars of different [Fe/H] have been plotted with different colors. Crosses and filled circles represent cluster and field stars, respectively. Error bars in bottom-right corner of each panel show the mean uncertainties.}
\label{fig_luminosity}
\end{figure*}

The profile of each CaT line is fitted by a combination of a Gaussian plus a Lorentzian. As was described in Paper I, this combination provides the best fit to the line core and wings for both weak and
strong lines. The fit is also good for the very weak lines of extremely metal-poor stars. This least square fit was performed using the Levenberg-Marquadt algorithm. Although in our case the spectra
had already been normalized, the position of the continuum was recalculated by
performing a linear fitting to the mean values of each continuum bandpass. Finally, the equivalent width is obtained as the area delimited by the 
profile fitted to the line and the continuum level. The equivalent
widths of each CaT line and their uncertainties determined for both field and cluster stars are
listed in Table~\ref{cluster_star_ew}.

Different ways of combining the strengths of the three
CaT lines can be found in the literature in order to obtain the global CaT index, $\Sigma Ca$ \citep[see][for a comparison among different index definitions]{cenarro2001}. The most used ones are the unweighted sum of the three features, $W_{8498}+W_{8542}+W_{8662}$
\citep[e.g.][Paper I]{cole2004}, or the unweighted sum of the two strongest
lines, $W_{8542}+W_{8662}$
\citep[e.g.][]{suntzeff1993,battaglia2008}. In the last one, the weakest line at
8498\AA~is excluded on the basis of its worse S/N.  In Fig.~\ref{fig_comp} we have plotted the run of $\Sigma Ca$, obtained as the sum of the equivalent widths of the three CaT lines,  versus the equivalent width of each CaT feature (panels a, b and c respectively) and the sum of the two strongest
lines (panel d). In all cases a clear linear correlation is observed. The contribution of the weakest line at 8498 \AA~to $\Sigma Ca$ is only the 19\%. The other two lines at 8542 and 8662 \AA~contribute with the 47\% and 34\%, respectively. Although we obtain our calibration as a function of $\Sigma Ca$, these relations allow to transform it to any other combination of the strength of the three CaT lines. We also investigated the ratio between the strength of the two strongest CaT lines. For all the stars in our sample we obtained $W_{8542}/W_{8662}$=1.32$\pm$0.09 without any dependence in absolute magnitude. Although within the dispersion, there is a small trend with metallicity in the sense that the ratio decreases with metallicity ($W_{8542}/W_{8662}=1.36+0.04\times[Fe/H]$. \citet{starkenburg2010} derived a mean ratio of 1.27 in their synthetic spectra. They also found a similar dependence with metallicity in their synthetic spectra. \citet{norris2008} obtained a ratio of 1.34 in globular cluster stars. Anyway, all these values are similar within the errorbars.

\section{The Ca\,{\sevensize\bf II} triplet metallicity scale}\label{sec6}

\subsection{A new CaT calibration valid for -4$\leq$[Fe/H]$\leq$+0.5}

\begin{table}
 \centering
\caption{Best fitting parameters\label{fittingparamters}}
  \begin{tabular}{@{}lcccc@{}}
\hline 
 & $V$ & $V-V_{HB}$ & $I$ & $K_S$\\
\hline
a & -3.45$\pm$0.04 & -3.45$\pm$0.04 & -3.43$\pm$0.04 & -3.33$\pm$0.05 \\
b & 0.16$\pm$0.01 & 0.11$\pm$0.02 & 0.13$\pm$0.01 & 0.15$\pm$0.01 \\
c & 0.41$\pm$0.004 & 0.44$\pm$0.006 & 0.45$\pm$0.006 & 0.48$\pm$0.008 \\
d & -0.53$\pm$0.11 & -0.65$\pm$0.12 & -0.50$\pm$0.12 & -0.27$\pm$0.13 \\
e & 0.019$\pm$0.002 & 0.03$\pm$0.003 & 0.016$\pm$0.002 & 0.01$\pm$0.002 \\
$\sigma$ & 0.17 & 0.16 & 0.16 & 0.17 \\
N & 422 & 401 & 321 & 413 \\
\hline
\end{tabular}
\end{table}

As explained above, the calibration of the CaT lines as metallicity indicator traditionally relies on the fact that stars of a given metallicity describe a linear sequence in the $\Sigma Ca$-Luminosity plane. Since the slope was supposed to be independent of metallicity, the variation of the zero-points with metallicity define the CaT calibration. This approximation fails for large luminosity ranges in the RGB (Paper I) and in particularly for extremely metal-poor stars \citep{starkenburg2010}. Moreover, \citet{starkenburg2010} noted that the zero-points of these sequences do not change linearly with metallicity as was assumed until that moment.

The run of $\Sigma Ca$ versus M$_I$, M$_K$, M$_V$, and $V-V_{HB}$ for the stars in our sample is shown in Fig.~\ref{fig_luminosity}. Stars of different metallicities have been plotted with different colors. From this figure, it is clearly noticed that the slopes and the shape of the sequences change as a function of metallicity for all four luminosity indicators used. This is more clear in the extremely metal-poor regime. The change of the shape of the sequences as a function of metallicity is also clearly noticed. To address these effects, \citet{starkenburg2010} proposed to add two new terms to the relationship used to obtain the CaT calibration: a cross term to account for the slope changes as a function of [Fe/H], and a term of $\Sigma Ca^{-1.5}$ to account for the changing offset. We have used the same analytic relationship to derive our calibration which is in the form:

\begin{equation}\label{ecu-1}
[Fe/H]=a+b\times Mag+c\times \Sigma Ca +d\times \Sigma Ca^{-1.5}+e\times\Sigma Ca\times Mag
\end{equation}

where $Mag$ refers to each luminosity indicator.

\begin{figure}
\includegraphics[width=85mm]{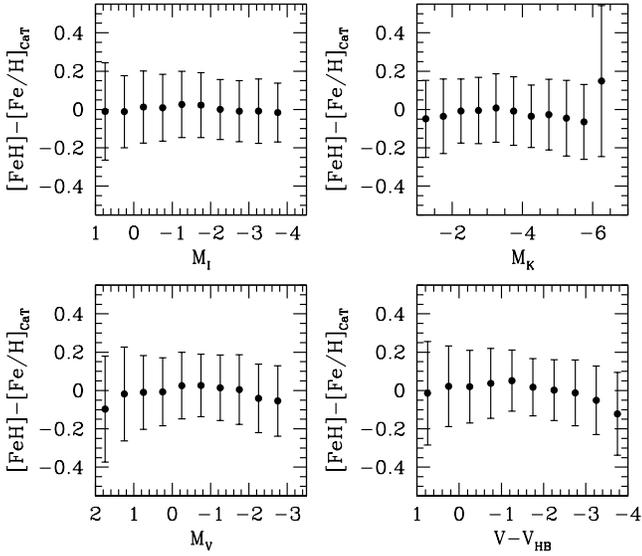}
\caption{Mean and sigma of the difference between the reference metallicities and those obtained from Equation~\ref{ecu-1} for each luminosity indicator.}
\label{fig_residuos}
\end{figure}

\begin{figure}
\includegraphics[width=85mm]{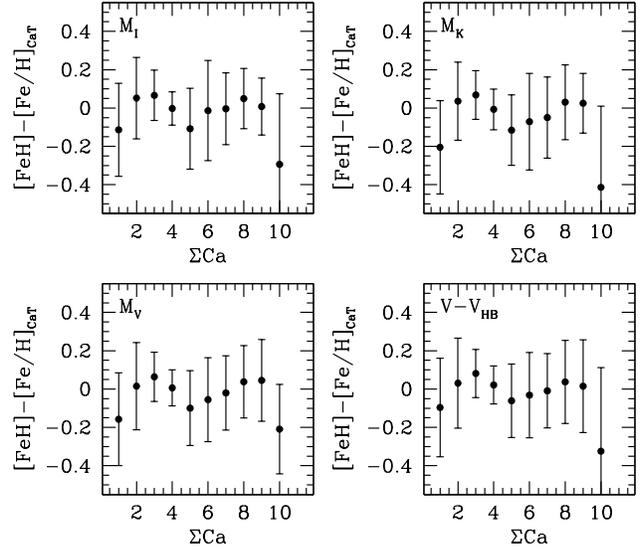}
\caption{As Fig.~\ref{fig_residuos} but as a function of $\Sigma Ca$.}
\label{fig_residuos2}
\end{figure}

The values obtained for the terms \textit{a}, \textit{b}, \textit{c}, \textit{d}, and \textit{e} for each luminosity indicator are listed in Table~\ref{fittingparamters} together with the number of stars used in each case and the sigma of the fit. The uncertainties of each parameter obtained by the normal least squares fit are very small. To compute more realistic uncertainties for each term we performed a Monte Carlo realisation. The input values of each star are randomly varied within their uncertainties assuming that they behave as a Gaussian probability distribution. A least square fit is performed using the modified input values. This procedure is repeated 5000 times. As expected, the values obtained as the mean of the results of each realisation are the same that those obtained by the original least square fit. The dispersion of these values provides a more realistic estimation of the uncertainty of each term which are also listed in Table~\ref{fittingparamters}. We have tested if the quality of the fit improve by changing the term $\Sigma Ca^{-1.5}$ for $\Sigma Ca^{2}$. However, the goodness of the fit worsens if the $\Sigma Ca^{2}$ term is used. We also investigated if the sigma of the fit decreases by adding quadratic terms. We found that adding new terms does not improve significantly the quality of the fit.

\begin{figure}
\includegraphics[width=85mm]{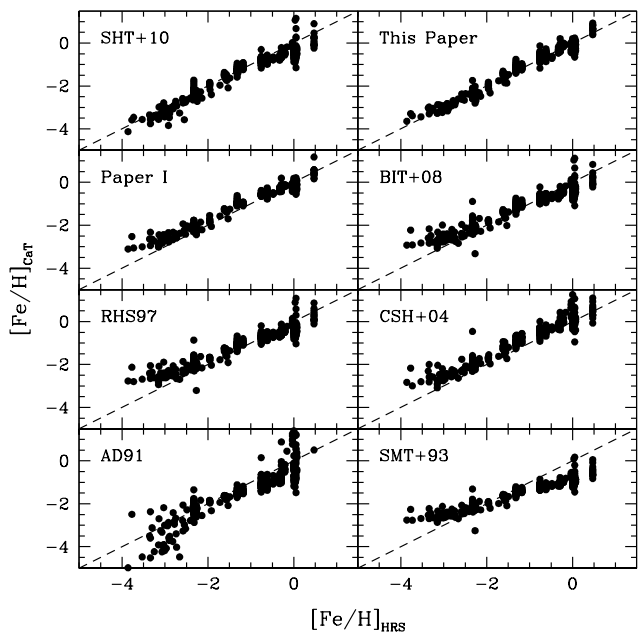}
\caption{Comparison among different CaT calibrations available in the literature and reference metallicities (see text and Table~\ref{table_comparison} for details).}
\label{fig_comparacion}
\end{figure}

The mean and sigma of the residuals of the fit as a function of each luminostiy indicator have been plotted in Fig.~\ref{fig_residuos} and Fig.~\ref{fig_residuos2}. Since there is not clear tendency neither with the position along the RGB nor with $\Sigma Ca$, and therefore with [Fe/H] we conclude that our calibration performs well for the whole range of metallicities studied. In general, the scatter of the residuals are above $\sim$0.15 dex which is slightly lower than the typical error bar of the [Fe/H] values derived form CaT calibrations which is $\sim$0.15--0.20 dex.

\subsection{Comparison with other calibrations}

\begin{table*}
 \centering
\caption{Compilation of the main features of the different CaT calibrations compared in this work.
\label{table_comparison}}
   \begin{tabular}{@{}lcccc@{}}
\hline
Source & $\Sigma Ca$ & Luminosity & Line fitting & Bandpasses\\
\hline
AD91 & W$_{8542}$+W$_{8662}$ & V-V$_{HB}$ & Gaussian & AD91 \\
SMT+93 & W$_{8542}$+W$_{8662}$ & V-V$_{HB}$ & Gaussian & AZ88 \\
RHS97 & 0.5W$_{8498}$+W$_{8542}$+0.6W$_{8662}$ & V-V$_{HB}$ & Moffat & RHS97 \\
CSH+04 & W$_{8498}$+W$_{8542}$+W$_{8662}$ & V-V$_{HB}$ & Gaussian+Lorentzian & AZ88 \\
Paper I & W$_{8498}$+W$_{8542}$+W$_{8662}$ & M$_{I}$ & Gaussian+Lorentzian & CCG+01 \\
BIT+08 & W$_{8542}$+W$_{8662}$ & V-V$_{HB}$ & Gaussian & BIT+08 \\
SHT+10 & W$_{8542}$+W$_{8662}$ & V-V$_{HB}$ & Gaussian & BIT+08 \\
This Paper & W$_{8498}$+W$_{8542}$+W$_{8662}$ & M$_{I}$ & Gaussian+Lorentzian & CCG+01 \\
\hline
\end{tabular}
\begin{minipage}{170mm}
References: AZ88: \citet{armandroffzinn1988}; AD91: \citet{armandroff_dacosta1991}; SMT+93: \citet{suntzeff1993}; RHS97: \citet{rutledge1997a,rutledge1997b}; CCG+01: \citet{cenarro2001}; BIT+08: \citet{battaglia2008}; SHT+10: \citet{starkenburg2010}.
\end{minipage}
\end{table*}

Several comparisons among the different CaT indices  can be found in the literature using stars in common among different studies \citep[][Paper I]{rutledge1997b}. However, to our knowledge, only \citet{battaglia2008} compared the final metallicities derived from their CaT calibration with reference values derived independently from high-resolution spectroscopy. It would be very useful to compare the metallicities derived from the different CaT calibrations available in the literature with the reference values obtained from the analysis of  both \mbox{Fe\,{\sc i}} and \mbox{Fe\,{\sc ii}} spectral lines in high-resolution spectra. The CaT calibrations are based on four ingredients: the bandpasses used to determine the continuum and spectral features; the way in which the line profile is fit; the reference metallicities used; and the form of the relationship among the CaT index, the luminosity indicator and the reference metallicities. In Section~\ref{sec1}, it has been explained in detail how different assumptions on some of these points affect the final CaT calibration.

We have selected eight reference works which are listed in Table~\ref{table_comparison}, together with the different approach used by each of them. They have been selected in order to include the most used bandpasses, line profile functions and relationships among CaT index, luminosity indicator and metallicity.

To perform this comparison we have determined the equivalent width of each CaT line in the same clusters and metal-poor stars used in this paper. We have followed the procedure that define each calibration, i.e. used the same bandpasses, line profile functions, index definitions, luminosity indicators and relationships among CaT index, luminosity indicator and reference metallicities. We refer the reader to the original papers for a detailed description of each procedure. 

\begin{figure}
\includegraphics[width=85mm]{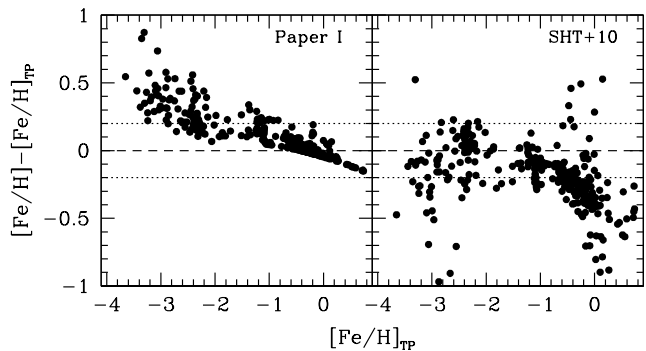}
\caption{Differences of the metallicities obtained by the calibration derived in Paper I (left) and those calculated with the relationship derived by \citet[][right]{starkenburg2010} with the values computed with the new calibration derived in this paper. Dotted lines represent a typical uncertainty of $\pm$0.2 dex.}
\label{fig_comparison2}
\end{figure}

In Fig.~\ref{fig_comparacion} we have plotted the high-resolution reference metallicities (X-axis) described in Section~\ref{sec4} versus the metallicity values obtained with each calibration (Y-axis). The \citet{armandroff_dacosta1991} calibration has the largest scatter, particularly at the metal-poor regime. This may be explained by the fact that it is the only one in which a quadratic relationship has been assumed between the reduced equivalent widths and metallicities. However, it behaves well in the metallicity range in which it was defined (-2$\leq$[Fe/H]$\leq$-0.7). All the calibrations before \citet{starkenburg2010}, including that obtained in Paper I, saturate at extremely metal-poor regimes ([Fe/H]$\leq$-2.5) as was noted by \citet{battaglia2008}. A large scatter is observed at metal-rich regimes ([Fe/H]$\geq$-0.25) with the exception of the one obtained in Paper I. These is observed even in the case of the one derived by \citet{cole2004} which used metal-rich open clusters as calibrators, together with metal-poor globular clusters. This denotes that together with the inclusion of the Lorentzian profile to sample the strong wings in metal-rich stars as was proposed by \citet{cole2004}, it is key to use bandpasses large enough to sample completely the wings of strong lines. The addition of two new terms in the relationship among the CaT index, luminosity indicator and metallicity proposed by \citet{starkenburg2010} solved the saturation problem at metal-poor metallicities. Although they adopted a correction procedure to account for the contribution of the damping wings of strong metal-rich to their single Gaussian fit, their calibration fails at metal-rich regimes. The calibration obtained in this paper seems to behave well in the whole range of metallicities covered.

In the left panel Figure~\ref{fig_comparison2} we have directly compared the metallicities calculated with the calibrations obtained in Paper I and in this paper. In general both calibrations produce similar metallicities within the uncertainties for -2$\leq$[Fe/H]$\leq$0. As expected, the calibration derived in paper I produces more metal-rich values for [Fe/H]$<$-2. This difference can be as large as 1 dex for [Fe/H]$<$-2.5. We perform the same comparison in right panel of Figure~\ref{fig_comparison2} with the metallicities computed from the calibration obtained by \citet[][]{starkenburg2010} from synthetic spectra. In general, there is a good agreement within the uncertainties between both calibrations for -3$\leq$[Fe/H]$\leq$-0.5. There are differences larger than 0.5 between the values obtained from both calibrations for metallicities more metal-rich than [Fe/H]$\sim$0. Again this is explained because their method do not sample properly the large wings of strong metal-rich lines. There are also large differences in the most metal-poor metallicities which may be explained by the difficulty of properly modelled the CaT lines at these extremely metal-poor metallicities.

\begin{figure}
\includegraphics[width=85mm]{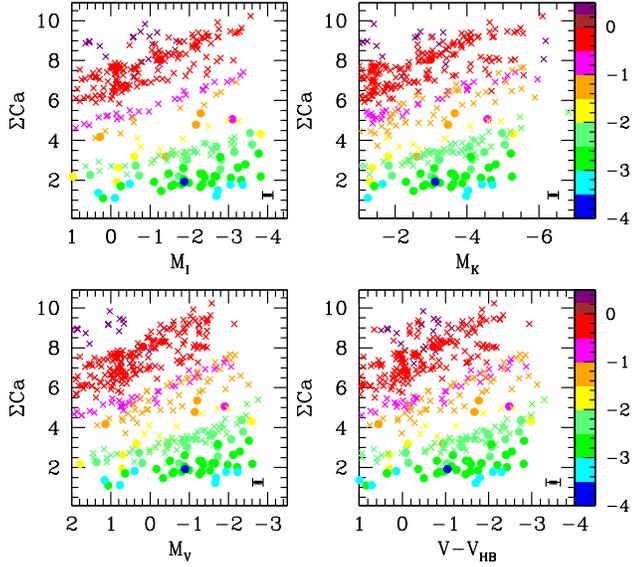}
\caption{As Fig.~\ref{fig_luminosity} but the colors are defined as a function of [Ca/H] abundances. }
\label{fig_luminosity_cah}
\end{figure}

\subsection{Dependence of the CaT lines with Ca abundances}

In spite that the strength of the CaT lines is expected to depend mainly on the Ca abundances rather than on the iron ones this is not the case, as it has been pointed by several investigations \citep[e.g.][]{idiart97,battaglia2008}. In fact, by comparing the Ca and Fe abundances determined from high-resolution spectroscopy and the CaT strength in medium-resolution spectra, \citet{battaglia2008} found that the CaT lines are a more robust estimator of [Fe/H] than of [Ca/H]. \citet{starkenburg2010} found that part of this discrepancy can be reduced if non-local thermodynamic equilibrium is taken into account to derive the calcium abundances from \mbox{Ca\,{\sc i}} lines. They suggest that the remaining discrepancy may relate to the outer atmospheric layers that are not very well modelled – even in non-local thermodynamic equilibrium. In this section we investigate whether the same behavior holds at the very low metallicity interval.

\begin{figure}
\includegraphics[width=84mm]{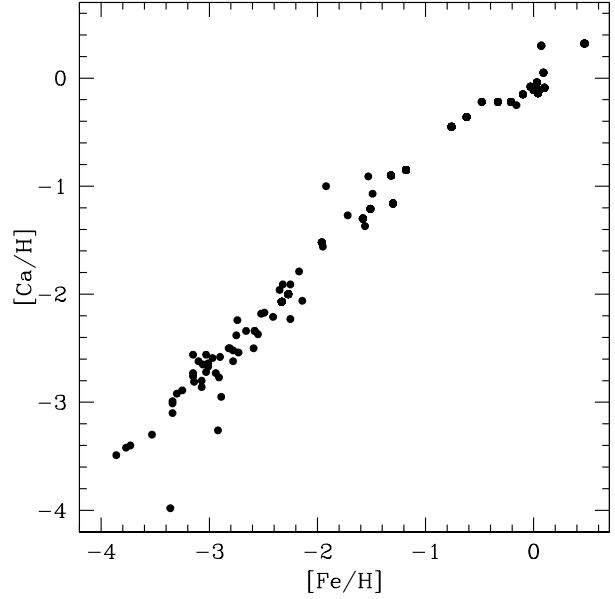}
\caption{Run of [Ca/H] versus [Fe/H] for the stars in our sample, down to the very low metallicities studied in this paper.}
\label{fig_cah}
\end{figure}

As in Fig.~\ref{fig_luminosity}, in Fig.~\ref{fig_luminosity_cah} we have plotted the run of $\Sigma Ca$ as a function of the different luminosity indicators used. In this case, the different colors denote different [Ca/H] ratios. As in the case of Fe, there is a clear trend with Ca abundances, as expected,  since [Fe/H] is correlated with  [Ca/H],  down to very low metallicities, as shown in Fig.~\ref{fig_cah}. However, comparison of Fig.~\ref{fig_luminosity}~and~\ref{fig_luminosity_cah}, reveals that $\Sigma$ Ca reflects more cleanly the Fe abundances than the Ca ones. As in Paper I, we conclude that to properly investigate the relation of the CaT equivalent widths with the Ca and Fe abundances, it may be necessary to sample objects with very different [Ca/H] ratios at a given [Fe/H].

\section{Summary}\label{sec7}

The cluster sample used in Paper I has been complemented with observations of extremely metal-poor field stars in order to obtain a new calibration of the strength of the infrared CaT lines as metallicity indicator. The obtained calibration is of very general use and applicability since:
i)  it is valid in the range -4$\leq$[Fe/H]$\leq$+0.5, which is the widest metallicity range in which the behavior of the CaT lines has been homogeneously investigated.
ii) it has been obtained on the basis of the traditional luminosity indicators V-V$_{HB}$, M$_V$, M$_I$, but also as a function of M$_K$. This will allow to use also the magnitudes derived by 2MASS, which covers almost the whole celestial sphere.
iii) it is valid for at least five magnitudes below the tip of the RGB.
iv) the contribution of each line to the global CaT index, obtained as $\Sigma Ca=W_{8498}+W_{8542}+W_{8662}$ has been determined. The strength of each line at 8498, 8542, and 8662 \AA\  contribute with the 19\%, 47\%, and 34\% to $\Sigma Ca$, respectively.

The calibration obtained in this paper results in a tight correlation between [Fe/H] abundances measured from high resolution spectra and [Fe/H] values derived from the CaT, over the whole metallicity range covered. Former saturations of the index, at low and high metallicities, which hampered the use of the CaT in these metallicity regimes, are no longer observed. We conclude, therefore, that the CaT remains a powerful metallicity indicator, applicable to nearby extragalactic star clusters and galaxies where high resolution metallicity measurements are not possible due to the faintness of the targets.

\section*{Acknowledgments}

We acknowledge the anonymous referee for comments and suggestions which have significantly improved the analysis and results presented in this paper. R.C. acknowledges funds provided by the Spanish Ministry of Science and Innovation under the Juan de la Cierva fellowship and under the Plan Nacional de Investigación Científica, Desarrollo, e Investigación Tecnolígica, AYA2010-16717. This research has made use of the WEBDA database, operated at the Institute for Astronomy of the University of Vienna, and the SIMBAD database, operated at CDS, Strasbourg, France

\bsp

\label{lastpage}

\end{document}